\documentclass[conference]{IEEEtran}

\usepackage[UKenglish]{babel}

\usepackage[paper=a4paper, left=2cm,right=2cm,top=2cm,bottom=2cm]{geometry}
\usepackage[hang,splitrule]{footmisc}
\usepackage[normalem]{ulem}
\usepackage[
pdftitle={Community Cloud Computing},
pdfauthor={Gerard Briscoe, Alexandros Marinos},
pdfsubject={Digital Ecosystems},
pdfkeywords={Cloud Computing, Community Clouds, Community Cloud Computing, Sustainability, Digital Ecosystems},
colorlinks=true, linkcolor=black, citecolor=black,  filecolor=black, urlcolor=black,
hyperfootnotes=false]{hyperref} 

\usepackage{graphicx, setspace, acronym, ifthen, substr, color}

\newboolean{final}\setboolean{final}{true}

\newboolean{mactex}\setboolean{mactex}{true}

\newboolean{arxiv}\setboolean{arxiv}{true}

\ifthenelse{\boolean{final}}{}{\usepackage{showlabels}\usepackage{citeext}}



\ifthenelse{\boolean{arxiv}}
	{\newcommand{\arxivfootnote}{\footnote}}
	{\def\arxivfootnote#1{}}

\ifthenelse{\boolean{arxiv}}
	{\def\arxivOnly#1{#1}}
	{\def\arxivOnly#1{}}

\ifthenelse{\boolean{arxiv}}
	{\def\arxivNot#1{}}
	{\def\arxivNot#1{#1}}

\ifthenelse{\boolean{arxiv}}
	{\def\arxiv#1#2{#1}}
	{\def\arxiv#1#2{#2}}

\newcommand{\tfigure}[9]
        {
        \IfSubStringInString{!}{#7}{\begin{figure}[#7]}{\begin{figure}[!t]}
        \IfSubStringInString{mm}{#8}{\vspace{#8}}{}
        \centering
        
        \IfSubStringInString{pdf}{#3}
                {
                \ifthenelse{\boolean{mactex}}{}{\execute{cd images; ln -s #2.pdf .#2.gdf}}
	      \includegraphics[#1]{images/#2}
                }
                {
                \ifthenelse{\boolean{mactex}}{}{\execute{cd images; ./pdfcrop.sh #2}}
                \includegraphics[#1]{images/#2-crop.pdf}
                }

        \vspace{#6}
        \caption[#4]
                {
                \label{#2}
                #4: #5
                }
        \IfSubStringInString{mm}{#9}{\vspace{#9}}{}
        \end{figure}
        }

\newcommand{\Circlesub}[4]
	{
	\ifthenelse{\boolean{mactex}}{}{\immediate\write18{cd images; ./pdfcrop.sh circle#2}}
	\ifthenelse{\boolean{final}}
		{\hspace{#1}\raisebox{#4}{$\includegraphics[scale=0.5, clip=true, trim=0mm 0mm 0mm 0mm]{images/circle#2-crop.pdf}$}\hspace{#3}}
		{\href{file://localhost/Users/g/Desktop/PhDthesis/images/circle#2.graffle}{\hspace{#1}\raisebox{#4}{$\includegraphics[clip=true, trim=0mm 0mm 0mm 0mm]{images/circle#2-crop.pdf}$}\hspace{#3}}}
	}

\newcommand{\execute}[1]{\immediate\write18{#1}}

\definecolor{tred}{RGB}{255,0,0}

\newcommand{\setCap}[2]{#1\immediate\write18{./mkcaption.sh #2}}
\newcommand{\getCap}[1]{\acl*{#1}}

\acrodef{PCG}{Projected Conjugate Gradient} 
\acrodef{QP}{quadratic programming}
\acrodef{RBF}{Radial-Basis Function}
\acrodef{ABM}{Agent-Based Modelling}
\acrodef{AI}{Artificial Intelligence}
\acrodef{DAI}{Distributed Artificial Intelligence}
\acrodef{API}{Application Programming Interface}
\acrodef{ARF}{p14ARF human tumor-suppressor gene}
\acrodef{B2B}{business-to-business}
\acrodef{BDP}{Biological Design Pattern}
\acrodef{BGS}{Best Guess Solution}
\acrodef{BIC}{Biologically-Inspired Computing}
\acrodef{BML}{Business Modelling Language}
\acrodef{BPEL}{Business Process Execution Language}
\acrodef{BPMN}{Business Process Modelling Notation}
\acrodef{CAS}{Complex Adaptive Systems}
\acrodef{COBOL}{COmmon Business-Oriented Language}
\acrodef{DBE}{Digital Business Ecosystem}
\acrodef{DE}{Digital Ecosystem}
\acrodef{DEC}{distributed evolutionary computing}
\acrodef{DGA}{Distributed genetic algorithms}
\acrodef{DIS}{Distributed Intelligence System}
\acrodef{DNA}{Deoxyribose Nucleic Acid}
\acrodef{DOP}{DBE Open Protocol}
\acrodef{DSS}{Distributed Storage System}
\acrodef{EAP}{Evolving Agent Population}
\acrodef{ebXML}{e-business eXtensible Markup Language}
\acrodef{EC}{Evolutionary Computing}
\acrodef{ECJ}{Evolutionary Computing in Java}
\acrodef{EE}{Evolutionary Environment}
\acrodef{EFL}{Evolutionary Framework for Language}
\acrodef{FLE}{Framework for Language Ecosystems}
\acrodef{EOA}{Ecosystem-Oriented Architecture}
\acrodef{ESS}{evolutionary stable strategy}
\acrodef{EvE}{Evolutionary Environment}
\acrodef{ExE}{Execution Environment}
\acrodef{FCB}{Framework for Computational Biomimicry}
\acrodef{FFF}{Fitness Function Framework}
\acrodef{FL}{Fitness Landscape}
\acrodef{HWU}{Heriot-Watt University}
\acrodef{ICL}{Imperial College London}
\acrodef{ICT}{Information and Communications Technology}
\acrodef{INTEL}{Intel Ireland}
\acrodef{IPA}{International Phonetic Alphabet}
\acrodef{ISUFI}{Istituto Superiore Universitario di Formazione Interdisciplinare}
\acrodef{JDJ}{Java Developer's Journal}
\acrodef{KC}{Kolmogorov-Chaitin}
\acrodef{LAN}{local area network}
\acrodef{LSE}{London School of Economics and Political Science}
\acrodef{MAS}{Multi-Agent System}
\acrodef{MDL}{Minimum Description Length}
\acrodef{MDM2}{murine double minute 2}
\acrodef{MFT}{Mean Field Theory}
\acrodef{MoAS}{Mobile Agent System}
\acrodef{MOF}{Meta Object Facility}
\acrodef{MUH}{migration and usage history}
\acrodef{NIC}{Nature Inspired Computing}
\acrodef{NN}{Neural Network}
\acrodef{NoE}{Network of Excellence}
\acrodef{OMG}{Open Mac Grid}
\acrodef{OPAALS}{Open Philosophies for Associative Autopoietic Digital Ecosystems}
\acrodef{P2P}{peer-to-peer}
\acrodef{P53}{protein 53}
\acrodef{PDA}{Personal Digital Assistant}
\acrodef{QoS}{quality of service}
\acrodef{REST}{REpresentational State Transfer}
\acrodef{RNA}{Deoxyribose Nucleic Acid}
\acrodef{SAE}{Software Agent Ecosystem}
\acrodef{SBML}{Systems Biology Modelling Language}
\acrodef{SBVR}{Semantics of Business Vocabulary and Business Rules}
\acrodef{SDL}{Service Description Language}
\acrodef{SF}{Service Factory}
\acrodef{SIM}{Social Interaction Mechanism}
\acrodef{SM}{Service Manifest}
\acrodef{SME}{Small and Medium sized Enterprise}
\acrodef{SML}{Service Modelling Language}
\acrodef{SMO}{Sequential Minimal Optimisation}
\acrodef{SOA}{Service-Oriented Architecture}
\acrodef{SOAP}{Simple Object Access Protocol}
\acrodef{SOC}{Self-Organised Criticality}
\acrodef{SOLUTA}{SOLUTA.NET}
\acrodef{SOM}{Self-Organising Map}
\acrodef{SSL}{Semantic Service Language}
\acrodef{STU}{Salzburg Technical University}
\acrodef{SUN}{Sun Microsystems}
\acrodef{SVM}{Support Vector Machine}
\acrodef{TM}{Turing Machine}
\acrodef{UBHAM}{University of Birmingham}
\acrodef{UDDI}{Universal Description Discovery and Integration}
\acrodef{UML}{Unified Modelling Language}
\acrodef{URI}{Uniform Resource Identifier}
\acrodef{UTM}{Universal Turing Machine}
\acrodef{VLP}{variable length population}
\acrodef{VLS}{variable length sequences}
\acrodef{vls}{variable length sequence}
\acrodef{WP}{Work-Package}
\acrodef{WSDL}{Web Services Definition Language}
\acrodef{XMI}{XML Metadata Interchange}
\acrodef{XML}{eXtensible Markup Language}
\acrodef{MD5}{Message-Digest algorithm 5}
\acrodef{GA}{genetic algorithm}
\acrodef{GP}{genetic programming}
\acrodef{MASON}{Multi-Agent Simulator Of Neighbourhoods}
\acrodef{Repast}{Recursive Porous Agent Simulation Toolkit}
\acrodef{JCLEC}{Java Computing Library for Evolutionary Computing}
\acrodef{OWL-S}{Web Ontology Language - Service}
\acrodef{EGT}{Evolutionary Game Theory}
\acrodef{RBF}{Radial Basis Functions}
\acrodef{SWS}{Semantic Web Services}
\acrodef{HDD}{Hard Disk Drive}
\acrodef{SSD}{Solid-State Drive}
\acrodef{OKS}{Open Knowledge Space}
\acrodef{CAES}{Complex Adaptive EcoSystem}

\acrodef{col3cap}{when consumers visit an application served by the central Cloud, which is housed in one or more data centres.}
\acrodef{colCap}{Green symbolises resource consumption, and yellow resource provision.}
\acrodef{col2Cap}{The role of coordinator for resource provision is designated by red, and is centrally controlled.}
\acrodef{absCap}{There is a significant \emph{buzz} \cite{buzz} around Cloud Computing, but there is little clarity about which offerings actually qualify and their interrelation. The key to resolving this confusion is by realising that the various offerings fall into different levels of abstraction,}
\acrodef{abs2Cap}{aimed at different market segments.}
\acrodef{c32cap}{shaping the underutilised resources of user machines}
\acrodef{c3cap}{with nodes potentially taking on all roles, \emph{consumer}, \emph{producer}, and most importantly \emph{coordinator}}

\newcommand{\white}[1]{\color{white}#1\normalcolor}

\begin{document}

\title{Digital Ecosystems in the Clouds:\\ Towards Community Cloud Computing}
\date{December 2008}
\author{
 \IEEEauthorblockN{Gerard Briscoe}
 \IEEEauthorblockA{Digital Ecosystems Lab\\
Department of Media and Communications\\
London School of Economics and Political Science\\
United Kingdom\\
e-mail: g.briscoe@lse.ac.uk}
\and 
 \IEEEauthorblockN{Alexandros Marinos}
 \IEEEauthorblockA{Department of Computing\\
Faculty of Engineering \& Physical Sciences\\
University of Surrey\\
United Kingdom\\
e-mail: a.marinos@surrey.ac.uk}
}
\maketitle

\begin{abstract}
\white{.} Cloud Computing is rising fast, with its data centres growing at an unprecedented rate. However, this has come with concerns of privacy, efficiency at the expense of resilience, and environmental sustainability, because of the dependence on Cloud vendors such as Google, Amazon, and Microsoft. Community Cloud Computing makes use of the principles of Digital Ecosystems to provide a paradigm for Clouds in the community, offering an alternative architecture for the use cases of Cloud Computing. It is more technically challenging to deal with issues of distributed computing, such as latency, differential resource management, and additional security requirements. However, these are not insurmountable challenges, and with the need to retain control over our digital lives and the potential environmental consequences, it is a challenge we must pursue.
\end{abstract}

\section{Introduction}
The recent development of Cloud Computing provides a compelling value proposition for organisations to outsource their \ac{ICT} infrastructure \cite{wiki1}. However, there are growing concerns over the control ceded to large Cloud vendors, including the lack of information privacy \cite{berkely}. Also, the data centres required for Cloud Computing are growing exponentially \cite{hayes}, creating an ever-increasing \emph{carbon footprint}, raising environmental concerns \cite{mckenna2008cws, mckinsey}.

The social paradigms and technologies of Digital Ecosystems, including the community ownership of digital infrastructure, can remedy these concerns. So, Cloud Computing combined with the principles of Digital Ecosystems provides a compelling socio-technical conceptualisation for sustainable distributed computing, utilising the spare resources of networked personal computers to provide the facilities of a virtual \emph{data centre} to form collectively a Community Cloud.

\section{Cloud Computing}

Cloud Computing is the use of Internet-based technologies for the provision of services \cite{wiki1}, originating from the \emph{cloud} as a metaphor for the Internet, based on how it is depicted in computer network diagrams to abstract the complex infrastructure it conceals \cite{wiki7}. It can be seen as a commercial evolution of the academia-oriented Grid Computing \cite{foster2008cca}, succeeding where Utility Computing struggled \cite{utilityComp, cellchip}. It is being promoted as the cutting edge of scalable web application development \cite{berkely}, in which dynamically scalable and often virtualised resources are provided as a service over the Internet \cite{wiki2, wiki1, wiki4, wiki5}, with users having no knowledge of, expertise in, or control over the technology infrastructure of the Cloud supporting them \cite{wiki6}. It currently has significant momentum in two extremes of the web development industry \cite{berkely, wiki1}: the consumer web technology incumbents who have resource surpluses in their vast \emph{data centres}\arxivfootnote{A \emph{data centre} is a facility, with the necessary security devices and environmental systems (e.g. air conditioning and fire suppression), for housing a \emph{server farm}, a collection of computer servers that can accomplish server needs far beyond the capability of one machine \cite{arregoces2003dcf}.}, and various consumers and start-ups that do not have access to such computational resources. Cloud Computing conceptually incorporates software-as-a-service (SaaS) \cite{saas}, Web 2.0 \cite{web2.0}  and other technologies with reliance on the Internet, providing common business applications online through web browsers to satisfy the computing needs of users, while the software and data are stored on the servers.

\tfigure{scale=\arxiv{2.0}{1.0}}{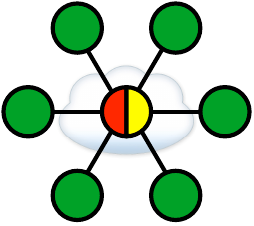}{graffle}{Cloud Computing}{Typical configuration \getCap{col3cap} \getCap{colCap} \getCap{col2Cap}}{\arxiv{-2}{-4}mm}{}{}{\arxivNot{-5mm}}

Figure \ref{cloudComputing} shows the typical configuration of Cloud Computing at run-time \setCap{when consumers visit an application served by the central Cloud, which is housed in one or more data centres.}{col3cap} \setCap{Green symbolises resource consumption, and yellow resource provision.}{colCap} \setCap{The role of coordinator for resource provision is designated by red, and is centrally controlled.}{col2Cap} From the figure, it can be seen that coordination and resource provision are centrally controlled, even if the central node is implemented as a distributed grid, which is the usual incarnation of a data centre. Providers, who are also the controllers, are usually companies with other web activities that require large computing resources, and in their efforts to scale their primary businesses they have gained considerable expertise and hardware. For them, Cloud Computing is a way to resell these as a new product while expanding into a new market. Consumers include everyday users, \acp{SME}, and ambitious start-ups whose innovation potentially threatens the incumbent providers.

\subsection{Layers of Abstraction}

\setCap{There is a significant \emph{buzz} \cite{buzz} around Cloud Computing, but there is little clarity about which offerings actually qualify and their interrelation. The key to resolving this confusion is by realising that the various offerings fall into different levels of abstraction,}{absCap} as shown in Figure \ref{Drawing1}, \setCap{aimed at different market segments.}{abs2Cap}

\subsubsection{Infrastructure as a Service (IaaS) \cite{iaas}} At the most basic level of the Cloud Computing offerings, there are providers such as Amazon \cite{amazon} and Mosso \cite{mosso}, who provide \emph{machine instances} to developers. These instances essentially behave like dedicated servers that are controlled by the developers, who therefore have full responsibility for their operation. So, once a machine reaches its performance limits, the developers have to manually instantiate another machine and scale their application out to it. This service is intended for developers who can write arbitrary software on top of the infrastructure with only small compromises in their development methodology.

\subsubsection{Platform as a Service (PaaS) \cite{paas}} One level of abstraction above, services like Google App Engine \cite{appEngine} provide a programming environment that abstracts machine instances and other technical details from developers. The programs are executed over data centres, not concerning the developers with matters of allocation. In exchange for this, the developers have to handle some constraints that the environment imposes on their application design, for example the use of \emph{key-value stores}\arxivfootnote{A distributed storage system for structured data that focuses on scalability, at the expense of the other benefits of relational databases \cite{keyvaluestore}, e.g. Google's BigTable \cite{bigtable} and Amazon's SimpleDB \cite{dynamo}.} instead of \emph{relational databases}.

\subsubsection{Software as a Service (SaaS) \cite{saas}} At the consumer-facing level are the most popular examples of Cloud Computing, with well-defined applications offering users online resources and storage. This differentiates SaaS from traditional websites or web applications which do not interface with user information (e.g. documents) or do so in a limited manner. Popular examples include Microsoft's (Windows Live) Hotmail, office suites such as Google Docs and Zoho, and online business software such as Salesforce.com.  

\tfigure{width=3.25in}{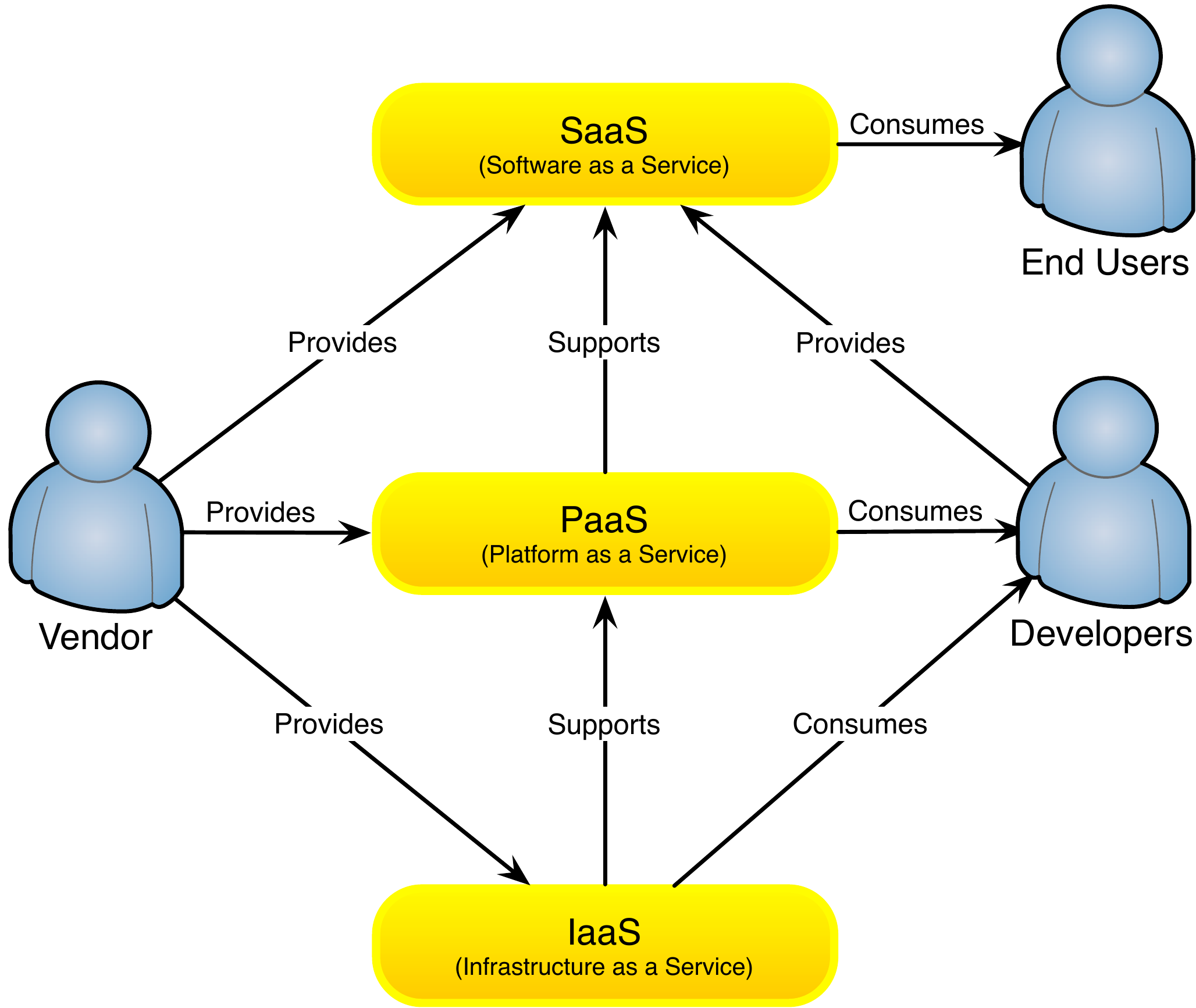}{graffle}{Abstractions of Cloud Computing}{\getCap{absCap} \getCap{abs2Cap}}{-6mm}{}{}{}

To better understand Cloud Computing we can categorise the roles of the various actors. The \emph{vendor} as resource provider has already been discussed. The application \emph{developers} utilise the resources provided, building services for the \emph{end users}. This separation of roles helps define the stakeholders and their differing interests. However, actors take on multiple roles, with \emph{vendors} developing services for the \emph{end users}, or \emph{developers} utilising the services of others for their own. Yet, within a Cloud the role of provider, and therefore controller, can only be occupied by a single entity, the \emph{vendor}. 
 
\subsection{Concerns}
The Cloud Computing model is not without concerns, as others have noted \cite{gnuMan, berkely}, and we consider the following as primary:

\subsubsection{Economics of Failure} The uptime\arxivfootnote{Uptime is a measure of the time a computer system has been running, i.e. up. It came into use to describe the opposite of downtime, times when a system was not operational \cite{mccabe2007naa}.}  of Cloud Computing-based solutions is an advantage, when compared to businesses running their own infrastructure, but often overlooked is the co-occurrence of downtime in vendor-driven \emph{monocultures}. The use of globally decentralised \emph{data centres} for vendor Clouds minimises failure, aiding its adoption. However, when these failures do occur it has a cascade effect, taking down organisations depending on the Cloud. This was illustrated by the Amazon (S3) Cloud outage \cite{amazonOutage}, which took with it several other dependent businesses. So, failures are now system-wide, instead of being partial or localised. Therefore, the efficiencies gained from centralising infrastructure for Cloud Computing will increasingly be at the expense of the Internet's resilience.

\subsubsection{Convenience vs Control} 
The growing popularity of Cloud Computing comes from its convenience, but also brings vendor control, an issue of ever-increasing concern. For example, Google Apps for in-house e-mail typically provides higher uptime \cite{montgomery2008}, but its failure \cite{perez2007} highlighted the issue of lock-in that comes from depending on vendor Clouds. The even greater concern is the loss of information privacy, with vendors having full access to the resources stored on their Clouds. In particularly sensitive cases of SMEs and start-ups, the provider-consumer relationship that Cloud Computing fosters between the owners of resources and their users could potentially be detrimental, as there is a conflict of interest for the providers. They profit by providing resources to up and coming players, but also wish to maintain dominant positions in their consumer-facing industries.

\subsubsection{Environmental Impact} The other major concern is the ever-increasing \emph{carbon footprint} from the \emph{exponential growth} \cite{hayes} of the data centres required for Cloud Computing. With the industry expected to exceed the airline industry by 2020 \cite{mckinsey}, this raises sustainability concerns \cite{mckenna2008cws}. The industry is being motivated to address the problem by legislation \cite{mckinsey, epaReport}, the operational limit of power grids (not being able to power any more servers) \cite{miller2006}, and the potential financial benefits \cite{mcIsaac2007, mckinsey}. Their primary solution is the use of \emph{virtualisation}\arxivfootnote{Virtualisation is the creation of a virtual version of a resource, such as a server, which can then be stored, migrated, duplicated, and instantiated as needed, improving scalability and work load management \cite{wolf2005vde}.} to maximise resource utilisation \cite{virtualisation}, but the problem remains \cite{brill2007icd, brodkin2008}.

While these issues are endemic to current Cloud Computing, they are not flaws in the Cloud conceptualisation, but of the vendor provision and implementation of Clouds \cite{appEngine, amazon}. There are attempts to address some of these concerns, such as avoiding vendor lock-in through a portability layer between different vendor Clouds, called a meta-Cloud or a Cloud-of-Clouds \cite{metaCloud}. While this would avoid vendor lock-in to an extent, it will not alleviate the other concerns raised. Also, there is an open source implementation of the Amazon (EC2) Cloud \cite{amazon}, called Eucalyptus \cite{nurmi2008eos}, which allows for a data centre to execute code compatible with Amazon's Cloud, providing \emph{private internal} Clouds. This would avoid vendor lock-in and provide information privacy, but only for those with their own data centres, and so is not really Cloud Computing (which by definition is to avoid needing one's own data centre \cite{wiki1}). So, vendor Clouds remain synonymous with Cloud Computing \cite{wiki2, wiki1, wiki4, wiki5}, while our response is an alternative model for the Cloud conceptualisation infused with the principles of Digital Ecosystems.

\section{Digital Ecosystems}

Digital Ecosystems are distributed adaptive open socio-technical systems, with properties of self-organisation, scalability and sustainability, inspired by natural ecosystems \cite{wikipediaDE}, and are emerging as a novel approach to the catalysis of sustainable regional development driven by \acp{SME} \cite{reden}. Digital Ecosystems aim to help local economic actors become active players in globalisation \cite{dini2008bid}, valorising their local culture and vocations, and enabling them to interact and create value networks \cite{dbebook} at the global level. Increasingly this approach, dubbed \emph{glocalisation}, is being considered a successful strategy of globalisation that preserves regional growth and identity \cite{robertson1994gog\arxivOnly{, swyngedouw1992mqg, khondker2004}}, and has been embraced by the mayors and decision-makers of thousands of municipalities \cite{glocalforum2004}. The community focused on the deployment of Digital Ecosystems, REgions for Digital Ecosystems Network (REDEN) \cite{reden}, is supported by projects such as the Digital Ecosystems Network of regions for (4) DissEmination and Knowledge Deployment (DEN4DEK) \cite{den4dek}, a thematic network that aims to share experiences and disseminate all the necessary knowledge that will allow regions to plan an effective deployment of Digital Ecosystems at all levels (economic, social, technical and political) to produce real impacts in the economic activities of European regions through the improvement of \ac{SME} business environments.

In a traditional market-based economy, made up of sellers and buyers, the parties exchange property \cite{delcloque2001dii}, while in a new network-based economy, made up of servers and clients, the parties share access to services and experiences \cite{delcloque2001dii}. Digital Ecosystems aim to support network-based economies reliant on next-generation \ac{ICT} that will extend the \ac{SOA} concept \cite{soa1w} with the automatic combining of available and applicable services in a scalable architecture, to meet business user requests for applications that facilitate business processes. Scalable resource provision has yet to be considered in Digital Ecosystems research. Without it Digital Ecosystems risk being subsumed into vendor Clouds at the infrastructure level, while striving for decentralisation at the service level, which would clearly be incompatible with its principles. So, the realisation of the Digital Ecosystems vision requires a form of Cloud Computing -- but within the principle of community-based infrastructure, where individual users share ownership \cite{bionetics}.

\section{Community Cloud}
Community Cloud Computing arises from concerns over the control of vendors in Cloud Computing and the observation that analogous concerns drive Digital Ecosystems research. It aspires to combine the principles of Digital Ecosystems with the use cases of Cloud Computing. Replacing vendor Clouds by \setCap{shaping the underutilised resources of user machines}{c32cap} to form a Community Cloud, \setCap{with nodes potentially taking on all roles, \emph{consumer}, \emph{producer}, and most importantly \emph{coordinator}}{c3cap}, as shown in Figure \ref{c3}.

\tfigure{scale=\arxiv{1.75}{1.0}}{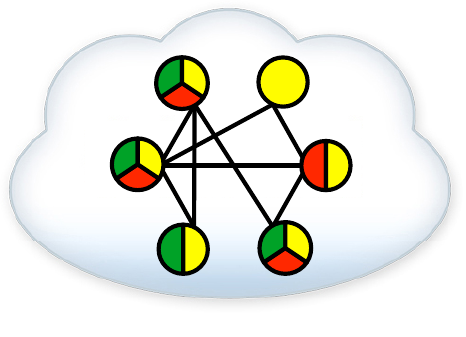}{graffle}{Community Cloud}{Created from \getCap{c32cap}, \getCap{c3cap}. Green symbolises resource consumption, yellow resource provision, and red resource coordination and administration.}{-12mm}{!b}{}{\arxivNot{-5mm}}

\subsection{Conceptualisation}
The conceptualisation of the Community Cloud draws from Cloud Computing \cite{wiki1}, Digital Ecosystems \cite{wikipediaDE} and Green Computing \cite{harris2008}. Its a paradigm for Cloud Computing in the \emph{community}, without dependence on Cloud vendors, such as Google, Amazon, or Microsoft.

\subsubsection{Openness} Removing the dependence on vendors makes the Community Cloud the open equivalent to vendor Clouds, and therefore identifies a new dimension in the open versus proprietary struggle \cite{west2001pvo} that has emerged in code, standards and data, but has not until now been expressed in the realm of hosted services.
 
\subsubsection{Community}
The Community Cloud is as much a social structure as a technology paradigm \cite{benkler2004sns}, because of the community ownership of the infrastructure. This community ownership carries with it a degree of economic scalability, without which there would be diminished competition and potential stifling of innovation as risked in vendor Clouds.

\subsubsection{Graceful Failures} The Community Cloud is not owned or controlled by any one organisation, and therefore not dependent on the lifespan or failure of any one organisation. It will be robust and resilient to failure, and immune to the system-wide cascade failures of vendor Clouds, because of the diversity of its supporting nodes. When occasionally failing it will do so gracefully, non-destructively, and with minimal downtime, as the unaffected nodes compensate for the failure.

\subsubsection{Convenience and Control} The Community Cloud, unlike vendor Clouds, has no inherent conflict between convenience and control, because its community ownership provides for democratic distributed control.

\subsubsection{Environmental Sustainability} The Community Cloud will have a significantly smaller \emph{carbon footprint} than vendor Clouds, because making use of underutilised user machines will require much less energy than the dedicated data centres required for vendor Clouds. The server farms within data centres are an intensive form of computing resource provision, while the Community Cloud is more organic, growing and shrinking in a symbiotic relationship to support the demands of the community, which in turn supports it.

\subsection{Architecture}

The method of materialising the Community Cloud is the distribution of its server functionality amongst a population of nodes provided by user machines, shaping their underutilised resources into a \emph{virtual data centre}. While straightforward in principle, this poses challenges on many different levels, but many are aligned with currently active research topics in Digital Ecosystems.

\subsubsection{Core Infrastructure}

Before proceeding to the resource exchange and service composition, the nodes will be deployed as isolated \emph{virtual machines}, forming a fully distributed peer-to-peer network, providing support for distributed identity and coordination.

\paragraph{Virtual Machines (VMs)}
Executing arbitrary code in the machine of a resource-providing user will require a \emph{sandbox}\arxivfootnote{A sandbox is a security mechanism for safely running programs, often used to execute untested code, or untrusted programs from unverified third-parties, suppliers and untrusted users \cite{sandbox}.} for the guest code, a VM\arxivfootnote{A virtual machine is a software implementation of a machine (computer) that executes programs like a real machine \cite{craig2006vm}.} to protect the host. The role of the VM is to make system resources \emph{safely} available to the Cloud in a measurable manner. Fortunately, feasibility has been shown with heavyweight VMs such as the Java Virtual Machine and Common Language Runtime, and with the lightweight JavaScript VMs present in most modern web browsers. Furthermore, the age \cite{geer2005cmt} of \emph{multi-core processors}\arxivfootnote{A multi-core processor is an integrated circuit to which two or more processors have been attached for enhanced performance, reduced power consumption, and more efficient simultaneous processing of multiple tasks \cite{zelkowitz2007}.} has resulted in unused and underutilised cores being commonplace in modern personal computers \cite{posey2007}, which lend themselves well to the deployment and background execution of Community Cloud facing VMs.

\paragraph{P2P Networking}
At this most fundamental level, nodes will have to be interconnected to form a peer-to-peer network. It will have to be specifically engineered to provide high resilience while avoiding single points of control and failure, which would make a decentralised super-peer based control mechanism \cite{risson2006srt} insufficient. A completely distributed peer-to-peer network is required, immune to super-peer failure.

\paragraph{Distributed Identity/Trust}
The performing of tasks beyond the networking requires nodes to identify each other and keep historical context. This identification must be performed in a fully distributed manner, which has implications as most identity schemes are based on an identity provider controlling provision. Additionally, trust should be tracked as a multi-dimensional variable, including considerations such as uptime, performance characteristics, and reputation. 

\subsubsection{Resource layer}

As the networking infrastructure is now in place, we can discuss the first consumer-facing uses of the \emph{virtual data centre} of the Community Cloud. At its core, Cloud Computing is about using resources from the Cloud. The Community Cloud will offer the usage experience of Cloud Computing on the platform-as-a-service (PaaS) layer and above. \emph{Utility Computing} \cite{rappa2004ubm} scenarios, such as access to raw storage and computation, will be made available at the PaaS layer. Access to these abstract resources for service deployment will then provide the software-as-a-service (SaaS) layer.

\paragraph{Distributed Computation}
The field has a long history of successful incarnations in its centrally controlled form. However, Community Cloud Computing will need to take inspiration from Grid Computing \cite{berman2003gcm} to provide distributed coordination of the computational capabilities that nodes offer to the Community Cloud. 

\paragraph{Distributed Persistence} 
The Community Cloud will naturally require storage on its participating nodes, taking advantage of the ever-increasing surplus on most\arxivfootnote{The only exception is the recent arrival of \ac{SSD} in personal devices, popular for mobile devices because of their lack of moving parts, and whose use is growing as their size and price reach traditional HDD \cite{ssd}.} personal computers \cite{diskTrend}. However, the method of information storage in the Community Cloud is an issue with multiple aspects. First, information can be file-based or structured. Second, while constant and instant availability can be crucial, there are scenarios in which recall times can be more relaxed. Such varying requirements call for a combination of approaches, including distributed storage \cite{yianilos2001efd}, distributed databases \cite{garciamolina2008dsc} and key-value stores \cite{keyvaluestore}. Information privacy in the Community Cloud will be provided by the encryption of user information when on remote nodes, only being unencrypted when cached on the user's node, allowing for the secure and distributed storage of information.

\paragraph{Bandwidth Management} 

The Community Cloud will probably require more bandwidth than vendor Clouds, but can take advantage of the ever-increasing bandwidth and deployment of broadband \cite{wallsten2008}. Also, peer-to-peer protocols such as BitTorrent have made the distribution of information over networks much less bandwidth-intensive for providers, accomplished by using the downloading peers as repeaters of the information they receive. Community Cloud Computing will have to adopt such approaches to ensure efficient use of available network bandwidth, to avoid fluctuations or sudden rises in demand (e.g. the Slashdot effect\arxivfootnote{The Slashdot effect, also known as slashdotting, is the phenomenon of a popular website linking to a smaller site, causing the smaller site to slow down or even temporarily close due to the increased traffic \cite{adler1999sea}.}) burdening parts of the network.

\paragraph{Community Currency}
An important theme in the Community Cloud is the notion of nodes being contributors as well as consumers, which will require a \emph{community currency}\footnote{In economics, a \emph{community currency} is a medium (currency) for exchanging goods and services within a community, that is not backed by a central authority (e.g. national government) \cite{greco2001mua}, and which need not be restricted geographically despite sometimes being called a local currency \cite{doteuchi2002cca}.} (redeemable against resources in the community) to reward users for offering valued resources. It will also allow for traditional Cloud vendors to participate, by offering their resources to the Community Cloud to gather considerable \emph{community currency}, which they can then monetise against participants who cannot contribute as much as they consume (i.e. running a \emph{community currency} deficit). To avoid predicting or hard-coding the relative cost of resources (storage, computation, bandwidth), their prices can fluctuate based on market demand.

\subsubsection{Service Layer}
Cloud Computing can be said to represent a move from service-oriented \cite{soa1w} to service-driven architectures, making services explicitly dependent on other providers instead of building on self-sufficient resource locations. Community Cloud Computing makes this more explicit, breaking down the stand-alone service paradigm, with any service by default being composed of resources contributed by multiple participants. The following sections describe the core infrastructural services that the Community Cloud needs to provide.

\paragraph{Distributed Service Repository}
The repository of the Community Cloud must provide persistence, as with traditional service repositories \cite{papazoglou2003soc}, for the pointers to services and the semantic descriptions of services. To support the absence of principal (service-producing) nodes during service execution, there must also be persistence of the executable code of services. Naturally, the implementation of a distributed service repository is made easier by the availability of the distributed storage infrastructure of the Community Cloud.

\paragraph{Remote Service Execution} 

When a service is needed to fulfil a request, but is not currently instantiated on a suitable node, a copy should be retrieved from the repository and instantiated as needed. This allows for flexible responsiveness and resilience to unpredictable traffic spikes. Nodes are naturally interested in executing services as their purpose is to gather \emph{community currency} for their users. A developer should note the resource cost of a service in its description, allowing for pre-execution resource budgeting by nodes, and post-execution \emph{community currency} payments by consumers. It is in a developer's own interest to mark resource costs correctly, because over-budgeting will burden their users and under-budgeting will cause premature service termination. Additionally, developers could add a subsidy to promote their services. Remote service execution must be secured against potentially compromised nodes, because while unable to access a complete traffic log of the services they execute, they could potentially access the business logic of the services they execute. Otherwise, we would be replacing the vendor \emph{looking in} problem, with an anyone \emph{looking in} problem.

\section{Wikipedia in the Community Cloud}

Wikipedia suffers from an ever-increasing demand for resources and bandwidth, without a stable revenue source for support \cite{heebie}. Their current funding model requires a continuous influx of monetary donations for the maintenance and expansion of their infrastructure \cite{regwikimon}, the alternative being contentious advertising revenues \cite{heebie}, which has caused a long-standing conflict within their community \cite{reuters}. While it would provide a more scalable funding model, the fear is it would compromise the public's trust in the content \cite{wikifear}. Alternatively, the Community Cloud could provide a self-sustaining scalable resource provision model without risk of compromising the content, because it would be compatible with their communal nature (unlike their current \emph{data centre} model), with their user base accomplishing the resource provision they require. 

Were Wikipedia to adopt Community Cloud Computing, it would be distributed throughout the Community Cloud alongside other services, which in this context can be as simple as a webpage or as complex as necessary. Participants in the Community Cloud will have a node on their machine, which when active accumulates \emph{community currency} by providing resources to fulfil service requests from other nodes. These service requests can be as simple as instantiating a simple HTML page or executing a server-side script, the core operations of Wikipedia. Participants can then use their amassed \emph{community currency} to interact with Wikipedia, performing a search or retrieving a page. More complicated tasks, such as editing a Wikipedia page, will require an update to the distributed storage of the Community Cloud, achieved by transmitting the new data through its network of nodes, most likely resulting in an eventual consistency model \cite{vogel}.

We have discussed Wikipedia in the Community Cloud, but the latter is not limited to not-for-profit organisations, being just as beneficial to for-profit businesses. Its organisational model for resource provision moves the cost of service provision to the user base, effectively creating a micro-payment scheme, which dramatically lowers the barrier of entry for innovative start-ups.

\section{Conclusions}

We have presented a socio-technical conceptualisation for sustainable distributed computing, the Community Cloud. The Community Cloud is an alternative to Cloud Computing, created from blending its usage scenarios with the principles of Digital Ecosystems. Community Cloud Computing utilises the spare resources of networked personal computers to provide the facilities of data centres, such that the community provides the computing power for the Cloud platform they wish to use. Furthermore, we hope that the Community Cloud will encourage innovation in vendor Clouds, forming a relationship analogous to the creative tension between open source and proprietary software.

\section*{Acknowledgements}
We would like to thank for helpful discussions, Georgios Exarchakos, Paul Krause, and Nick Antonopoulos of the university of Surrey, and Jo Stanley of the University of Cambridge. This work was supported by the EU-funded \ac{OPAALS} Network of Excellence (NoE), Contract No. FP6/IST-034824.

\execute{./urlfix.sh c3.bbl}
\bibliographystyle{IEEEtran.bst}

\nocite{hewitt2008osr}
\bibliography{references,cloudReferences}

\begin{thebibliography}{10}
\providecommand{\url}[1]{#1}
\csname url@samestyle\endcsname
\providecommand{\newblock}{\relax}
\providecommand{\bibinfo}[2]{#2}
\providecommand{\BIBentrySTDinterwordspacing}{\spaceskip=0pt\relax}
\providecommand{\BIBentryALTinterwordstretchfactor}{4}
\providecommand{\BIBentryALTinterwordspacing}{\spaceskip=\fontdimen2\font plus
\BIBentryALTinterwordstretchfactor\fontdimen3\font minus
  \fontdimen4\font\relax}
\providecommand{\BIBforeignlanguage}[2]{{\expandafter\ifx\csname l@#1\endcsname\relax
\typeout{** WARNING: IEEEtran.bst: No hyphenation pattern has been}\typeout{** loaded for the language `#1'. Using the pattern for}\typeout{** the default language instead.}\else
\language=\csname l@#1\endcsname
\fi
#2}}
\providecommand{\BIBdecl}{\relax}
\BIBdecl

\bibitem{wiki1}
M.~Haynie, ``Enterprise cloud services: Deriving business value from {C}loud
  {C}omputing,'' Micro Focus, Tech. Rep., 2009.

\bibitem{berkely}
\BIBentryALTinterwordspacing
M.~Armbrust, A.~Fox, R.~Griffith, A.~Joseph, R.~Katz, A.~Konwinski, G.~Lee,
  D.~Patterson, A.~Rabkin, I.~Stoica, and M.~Zaharia, ``Above the {C}louds: {A}
  {B}erkeley view of {C}loud {C}omputing,'' University of California, Berkeley,
  Tech. Rep., 2009. [Online]. Available:
  \url{http://d1smfj0g31qzek.cloudfront.net/abovetheclouds.pdf}
\BIBentrySTDinterwordspacing

\bibitem{hayes}
\BIBentryALTinterwordspacing
J.~Hayes, ``Cred - or croak?'' IET Knowledge Network, Tech. Rep., 2008.
  [Online]. Available:
  \url{http://kn.theiet.org/magazine/issues/0820/cred-croak-0820.cfm?SaveToPDF}
\BIBentrySTDinterwordspacing

\bibitem{mckenna2008cws}
P.~Mckenna, ``Can we stop the internet destroying our planet?'' \emph{New
  Scientist}, vol. 197, no. 2637, pp. 20--21, 2008.

\bibitem{mckinsey}
\BIBentryALTinterwordspacing
J.~Kaplan, W.~Forrest, and N.~Kindler, ``Revolutionizing data center energy
  efficiency,'' McKinsey \& Company, Tech. Rep., 2008. [Online]. Available:
  \url{http://www.mckinsey.com/clientservice/bto/pointofview/pdf/Revolutionizing_Data_Center_Efficiency.pdf}
\BIBentrySTDinterwordspacing

\bibitem{wiki7}
\BIBentryALTinterwordspacing
J.~Scanlon and B.~Wieners. (1999) The internet cloud. [Online]. Available:
  \url{http://www.thestandard.com/article/0,1902,5466,00.html}
\BIBentrySTDinterwordspacing

\bibitem{foster2008cca}
I.~Foster, Y.~Zhao, I.~Raicu, and S.~Lu, ``Cloud {C}omputing and {G}rid
  {C}omputing 360-degree compared,'' in \emph{Grid Computing Environments
  Workshop}, 2008, pp. 1--10.

\bibitem{utilityComp}
\BIBentryALTinterwordspacing
T.~Foremski. (2006) Sun services {CTO} says utility computing acceptance is
  slow going. [Online]. Available: \url{http://blogs.zdnet.com/Foremski/?p=33}
\BIBentrySTDinterwordspacing

\bibitem{cellchip}
\BIBentryALTinterwordspacing
A.~Orlowski. (2005) The {Cell} chip - how will {MS} and {I}ntel face the music?
  [Online]. Available:
  \url{http://www.theregister.co.uk/2005/02/03/cell_analysis_part_two/}
\BIBentrySTDinterwordspacing

\bibitem{wiki2}
\BIBentryALTinterwordspacing
G.~Gruman and E.~Knorr. (2008) What {C}loud {C}omputing really means. [Online].
  Available:
  \url{http://www.infoworld.com/article/08/04/07/15FE-cloud-computing-reality_1.html}
\BIBentrySTDinterwordspacing

\bibitem{wiki4}
\BIBentryALTinterwordspacing
Gartner. (2008) {C}loud {C}omputing will be as influential as e-business.
  [Online]. Available: \url{http://www.gartner.com/it/page.jsp?id=707508}
\BIBentrySTDinterwordspacing

\bibitem{wiki5}
\BIBentryALTinterwordspacing
P.~Gaw. (2008) What's the difference between {C}loud {C}omputing and {SaaS}?
  [Online]. Available:
  \url{http://blog.fortiva.com/fortivablog/2008/05/what-is-the-dif.html}
\BIBentrySTDinterwordspacing

\bibitem{wiki6}
\BIBentryALTinterwordspacing
K.~Danielson. (2008) Distinguishing {C}loud {C}omputing from {U}tility
  {C}omputing. [Online]. Available:
  \url{http://www.ebizq.net/blogs/saasweek/2008/03/distinguishing_cloud_computing/}
\BIBentrySTDinterwordspacing

\bibitem{arregoces2003dcf}
M.~Arregoces and M.~Portolani, \emph{Data center fundamentals}.\hskip 1em plus
  0.5em minus 0.4em\relax Cisco Press, 2003.

\bibitem{saas}
M.~Turner, D.~Budgen, and P.~Brereton, ``Turning software into a service,''
  \emph{Computer}, vol.~36, no.~10, pp. 38--44, 2003.

\bibitem{web2.0}
\BIBentryALTinterwordspacing
T.~Oreilly, ``What is {W}eb 2.0: {D}esign patterns and business models for the
  next generation of software,'' O'Reilly Media, Tech. Rep., 2008. [Online].
  Available:
  \url{http://www.oreillynet.com/pub/a/oreilly/tim/news/2005/09/30/what-is-web-20.html}
\BIBentrySTDinterwordspacing

\bibitem{buzz}
\BIBentryALTinterwordspacing
B.~Worthen. (2008) Overuse clouds buzz term's meaning. [Online]. Available:
  \url{http://blogs.wsj.com/biztech/2008/09/23/overuse-clouds-buzz-terms-meaning/}
\BIBentrySTDinterwordspacing

\bibitem{iaas}
A.~Newman, A.~Steinberg, and J.~Thomas, \emph{Enterprise 2.0
  Implementation}.\hskip 1em plus 0.5em minus 0.4em\relax McGraw-Hill Osborne
  Media, 2008.

\bibitem{amazon}
\BIBentryALTinterwordspacing
Amazon. (2009) Amazon {E}lastic {C}ompute {C}loud ({EC2}). [Online]. Available:
  \url{http://aws.amazon.com/ec2/}
\BIBentrySTDinterwordspacing

\bibitem{mosso}
\BIBentryALTinterwordspacing
Mosso. (2009) Deploy and scale websites, servers and storage in minutes.
  [Online]. Available: \url{http://www.mosso.com/}
\BIBentrySTDinterwordspacing

\bibitem{paas}
R.~Buyya, C.~Yeo, and S.~Venugopal, ``Market-oriented {C}loud {C}omputing:
  {V}ision, hype, and reality for delivering it services as computing
  utilities,'' in \emph{High Performance Computing and Communications}.\hskip
  1em plus 0.5em minus 0.4em\relax IEEE Press, 2008.

\bibitem{appEngine}
\BIBentryALTinterwordspacing
Google. (2009) Google {A}pp {E}ngine: {R}un your web apps on {G}oogle's
  infrastructure. [Online]. Available: \url{http://code.google.com/appengine/}
\BIBentrySTDinterwordspacing

\bibitem{keyvaluestore}
\BIBentryALTinterwordspacing
T.~Bain. (2008) Is the relational database doomed? [Online]. Available:
  \url{http://www.readwriteweb.com/archives/is_the_relational_database_doomed.php}
\BIBentrySTDinterwordspacing

\bibitem{bigtable}
F.~Chang, J.~Dean, S.~Ghemawat, W.~Hsieh, D.~Wallach, M.~Burrows, T.~Chandra,
  A.~Fikes, and R.~Gruber, ``Bigtable: A distributed storage system for
  structured data,'' in \emph{USENIX Symposium on Operating Systems Design and
  Implementation}, 2006.

\bibitem{dynamo}
G.~DeCandia, D.~Hastorun, M.~Jampani, G.~Kakulapati, A.~Lakshman, A.~Pilchin,
  S.~Sivasubramanian, P.~Vosshall, and W.~Vogels, ``Dynamo: {A}mazon's highly
  available key-value store,'' in \emph{Symposium on Operating Systems
  Principles}.\hskip 1em plus 0.5em minus 0.4em\relax ACM, 2007, pp. 205--220.

\bibitem{gnuMan}
\BIBentryALTinterwordspacing
B.~Johnson. (2008) Cloud {C}omputing is a trap, warns {GNU} founder {R}ichard
  {S}tallman. [Online]. Available:
  \url{http://www.guardian.co.uk/technology/2008/sep/29/cloud.computing.richard.stallman}
\BIBentrySTDinterwordspacing

\bibitem{mccabe2007naa}
\BIBentryALTinterwordspacing
J.~McCabe, \emph{Network analysis, architecture, and design}.\hskip 1em plus
  0.5em minus 0.4em\relax Morgan Kaufmann, 2007. [Online]. Available:
  \url{http://books.google.co.uk/books?id=iddGPgR48_MC}
\BIBentrySTDinterwordspacing

\bibitem{amazonOutage}
\BIBentryALTinterwordspacing
A.~Modine. (2008) Web startups crumble under {A}mazon {S}3 outage. [Online].
  Available:
  \url{http://www.theregister.co.uk/2008/02/15/amazon_s3_outage_feb_2008/}
\BIBentrySTDinterwordspacing

\bibitem{montgomery2008}
\BIBentryALTinterwordspacing
J.~Montgomery. (2008) Google {A}pps sees 99.9\%\ uptime, proves ``cloud
  reliability''. [Online]. Available:
  \\ \url{http://tech.blorge.com/Structure:0/2008/11/02/google-apps-sees-999-uptime-proves-cloud-reliability/}
\BIBentrySTDinterwordspacing

\bibitem{perez2007}
\BIBentryALTinterwordspacing
J.~Perez. (2007) Google {A}pps customers miffed over downtime. [Online].
  Available:
  \\ \url{http://www.pcworld.com/businesscenter/article/130234/google_apps_customers_miffed_over_downtime.html}
\BIBentrySTDinterwordspacing

\bibitem{epaReport}
``{EPA} report to congress on server and data center energy efficiency,'' US
  Congress, Tech. Rep., 2007.

\bibitem{miller2006}
\BIBentryALTinterwordspacing
R.~Miller. (2006) {NSA} maxes out {B}altimore power grid. [Online]. Available:
  \url{http://www.datacenterknowledge.com/archives/2006/08/06/nsa-maxes-out-baltimore-power-grid/}
\BIBentrySTDinterwordspacing

\bibitem{mcIsaac2007}
K.~McIsaac, ``The data centre goes green, the {CFO} saves money,'' Intelligent
  Business Research Services, Tech. Rep., 2007.

\bibitem{wolf2005vde}
C.~Wolf and E.~Halter, \emph{Virtualization: from the desktop to the
  enterprise}.\hskip 1em plus 0.5em minus 0.4em\relax Apress, 2005.

\bibitem{virtualisation}
R.~Talaber, T.~Brey, and L.~Lamers, ``Using virtualization to improve data
  center efficiency,'' The Green Grid, Tech. Rep., 2009.

\bibitem{brill2007icd}
K.~Brill, ``The invisible crisis in the data center: The economic meltdown of
  {M}oore's law,'' Uptime Institute, Tech. Rep., 2007.

\bibitem{brodkin2008}
\BIBentryALTinterwordspacing
J.~Brodkin. (2008) Gartner in `green' data centre warning. [Online]. Available:
  \url{http://www.techworld.com/green-it/news/index.cfm?newsid=106292}
\BIBentrySTDinterwordspacing

\bibitem{metaCloud}
\BIBentryALTinterwordspacing
C.~Metz. (2009) The {M}eta {C}loud - flying data centers enter fourth
  dimension. [Online]. Available:
  \url{http://www.theregister.co.uk/2009/02/24/the_meta_cloud/}
\BIBentrySTDinterwordspacing

\bibitem{nurmi2008eos}
D.~Nurmi, R.~Wolski, C.~Grzegorczyk, G.~Obertelli, S.~Soman, L.~Youseff, and
  D.~Zagorodnov, ``The {E}ucalyptus open-source cloud-computing system,'' in
  \emph{Cloud Computing and Its Applications}, 2008.

\bibitem{wikipediaDE}
\BIBentryALTinterwordspacing
Wikipedia. Digital {E}cosystem. [Online]. Available:
  \url{http://en.wikipedia.org/wiki/Digital_ecosystem}
\BIBentrySTDinterwordspacing

\bibitem{reden}
\BIBentryALTinterwordspacing
L.~{Rivera Le{\'o}n}. Regions for {D}igital {E}cosystems {N}etwork ({REDEN}).
  [Online]. Available: \url{http://reden.opaals.org/}
\BIBentrySTDinterwordspacing

\bibitem{dini2008bid}
P.~Dini, G.~Lombardo, R.~Mansell, A.~Razavi, S.~Moschoyiannis, P.~Krause,
  A.~Nicolai, and L.~Rivera~Le{\'o}n, ``Beyond interoperability to digital
  ecosystems: regional innovation and socio-economic development led by
  {SME}s,'' \emph{International Journal of Technological Learning, Innovation
  and Development}, vol.~1, pp. 410--426, 2008.

\bibitem{dbebook}
F.~Nachira, A.~Nicolai, P.~Dini, M.~Le~Louarn, and L.~Rivera~Le{\'o}n, Eds.,
  \emph{Digital {B}usiness {E}cosystems}.\hskip 1em plus 0.5em minus
  0.4em\relax European {C}ommission, 2007.

\bibitem{robertson1994gog}
R.~Robertson, ``Globalisation or glocalisation,'' \emph{Journal of
  International Communication}, vol.~1, pp. 33--52, 1994.

\bibitem{swyngedouw1992mqg}
E.~Swyngedouw, ``The mammon quest. `{G}localisation', interspatial competition
  and the monetary order: the construction of new scales,'' in \emph{Cities and
  regions in the new Europe: The Global-local Interplay and Spatial Development
  Strategies}, M.~Dunford and G.~Kafkalas, Eds.\hskip 1em plus 0.5em minus
  0.4em\relax Wiley, 1992, pp. 39--67.

\bibitem{khondker2004}
H.~Khondker, ``Glocalization as globalization: Evolution of a sociological
  concept,'' \emph{Bangladesh e-Journal of Sociology}, vol.~1, 2004.

\bibitem{glocalforum2004}
\BIBentryALTinterwordspacing
{Glocal Forum} and {CERFE}, ``The glocalization manifesto,'' The Glocal Forum,
  Tech. Rep., 2004. [Online]. Available:
  \url{http://www.glocalforum.org/mediagallery/mediaDownload.php?mm=/warehouse/documents/the_glocalization_manifesto.pdf}
\BIBentrySTDinterwordspacing

\bibitem{den4dek}
\BIBentryALTinterwordspacing
L.~{Rivera Le{\'o}n}. Digital {E}cosystems {N}etwork of {R}egions for
  {D}issemination and {K}nowledge {D}eployment ({DEN4DEK}). [Online].
  Available: \url{http://www.den4dek.org}
\BIBentrySTDinterwordspacing

\bibitem{delcloque2001dii}
P.~Delcloque and A.~Bramoull{\'e}, ``{DISSEMINATE}, an initial implementation
  proposal: a new point of departue in call for the `year 01'?'' \emph{ReCALL},
  vol.~13, pp. 277--292, 2001.

\bibitem{soa1w}
E.~Newcomer and G.~Lomow, \emph{Understanding {SOA} with web services}.\hskip
  1em plus 0.5em minus 0.4em\relax Addison-Wesley, 2005.

\bibitem{bionetics}
\BIBentryALTinterwordspacing
G.~Briscoe and P.~{De Wilde}, ``Digital {E}cosystems: Evolving service-oriented
  architectures,'' in \emph{Conference on Bio Inspired Models of Network,
  Information and Computing Systems}.\hskip 1em plus 0.5em minus 0.4em\relax
  IEEE Press, 2006. [Online]. Available: \url{http://arxiv.org/abs/0712.4102}
\BIBentrySTDinterwordspacing

\bibitem{harris2008}
J.~Harris, \emph{Green Computing and Green IT Best Practices}.\hskip 1em plus
  0.5em minus 0.4em\relax Lulu.com, 2008.

\bibitem{west2001pvo}
J.~West and J.~Dedrick, ``Proprietary vs. open standards in the network era:
  {A}n examination of the linux phenomenon,'' in \emph{System Sciences}.\hskip
  1em plus 0.5em minus 0.4em\relax IEEE Press, 2001, p.~10.

\bibitem{benkler2004sns}
Y.~Benkler, ``Sharing nicely: on shareable goods and the emergence of sharing
  as a modality of economic production,'' \emph{The Yale Law Journal}, vol.
  114, no.~2, pp. 273--359, 2004.

\bibitem{sandbox}
M.~Bishop, \emph{Computer Security}.\hskip 1em plus 0.5em minus 0.4em\relax
  Addison-Wesley, 2004.

\bibitem{craig2006vm}
I.~Craig, \emph{Virtual machines}.\hskip 1em plus 0.5em minus 0.4em\relax
  Springer, 2006.

\bibitem{geer2005cmt}
D.~Geer, ``Chip makers turn to multicore processors,'' \emph{IEEE Computer},
  vol.~38, no.~5, pp. 11--13, 2005.

\bibitem{zelkowitz2007}
M.~Zelkowitz, \emph{Advances in Computers: Architectural Issues}.\hskip 1em
  plus 0.5em minus 0.4em\relax Academic Press, 2007.

\bibitem{posey2007}
\BIBentryALTinterwordspacing
B.~Posey. (2007) Multi-core processors: Their implication for {W}indows.
  [Online]. Available:
  \url{http://searchwindowsserver.techtarget.com/tip/0,289483,sid68_gci1248527,00.html}
\BIBentrySTDinterwordspacing

\bibitem{risson2006srt}
J.~Risson and T.~Moors, ``Survey of research towards robust peer-to-peer
  networks: Search methods,'' \emph{Computer Networks}, 2006.

\bibitem{rappa2004ubm}
M.~Rappa, ``The utility business model and the future of computing services,''
  \emph{IBM Systems Journal}, vol.~43, no.~1, pp. 32--42, 2004.

\bibitem{berman2003gcm}
F.~Berman, G.~Fox, and A.~Hey, \emph{Grid {C}omputing: {M}aking the global
  infrastructure a reality}.\hskip 1em plus 0.5em minus 0.4em\relax Wiley,
  2003.

\bibitem{ssd}
\BIBentryALTinterwordspacing
C.~Mellor, ``Ssd and hdd capacity goes on embiggening,'' The Register, Tech.
  Rep., 2009. [Online]. Available:
  \url{http://www.theregister.co.uk/2009/01/09/ssd_and_hdd_capacity_increases/}
\BIBentrySTDinterwordspacing

\bibitem{diskTrend}
\BIBentryALTinterwordspacing
M.~Daley. (2009) Software bloat. [Online]. Available:
  \url{http://www.mattscomputertrends.com/softwarebloat.html}
\BIBentrySTDinterwordspacing

\bibitem{yianilos2001efd}
P.~Yianilos and S.~Sobti, ``The evolving field of distributed storage,''
  \emph{IEEE Internet Computing}, vol.~5, pp. 35--39, 2001.

\bibitem{garciamolina2008dsc}
H.~Garcia-Molina, J.~Ullman, and J.~Widom, \emph{Database {S}ystems: {T}he
  complete book}.\hskip 1em plus 0.5em minus 0.4em\relax Prentice Hall, 2008.

\bibitem{wallsten2008}
``Broadband growth and policies in {OECD} countries,'' Organisation for
  Economic Co-operation and Development, Tech. Rep., 2008.

\bibitem{adler1999sea}
S.~Adler, ``The {S}lashdot effect: {A}n analysis of three internet
  publications,'' \emph{Linux Gazette}, vol.~38, 1999.

\bibitem{greco2001mua}
T.~Greco, \emph{Money: {U}nderstanding and creating alternatives to legal
  tender}.\hskip 1em plus 0.5em minus 0.4em\relax Chelsea Green, 2001.

\bibitem{doteuchi2002cca}
\BIBentryALTinterwordspacing
A.~Doteuchi, ``Community currency and {NPO}s- {A} model for solving social
  issues in the 21st century,'' Social Development Research Group, NLI
  Research, Tech. Rep., 2002. [Online]. Available:
  \url{http://www.nli-research.co.jp/english/socioeconomics/2002/li0204a.pdf}
\BIBentrySTDinterwordspacing

\bibitem{papazoglou2003soc}
M.~Papazoglou, ``Service-oriented computing: concepts, characteristics and
  directions,'' in \emph{International Conference on Web Information Systems
  Engineering}, T.~Catarci, M.~Mecella, J.~Mylopoulos, and M.~Orlowska,
  Eds.\hskip 1em plus 0.5em minus 0.4em\relax IEEE Press, 2003, pp. 3--12.

\bibitem{heebie}
\BIBentryALTinterwordspacing
{Heebie Blog}. (2009) Wikipedia {F}undraising: {T}he real truth. [Online].
  Available: \\ \url{http://blog.heebie.co.uk/wikipedia-fundraising-real-truth}
\BIBentrySTDinterwordspacing

\bibitem{regwikimon}
\BIBentryALTinterwordspacing
A.~Modine. (2009) Wales' personal begging earns last \$2m. [Online]. Available:
  \url{http://www.theregister.co.uk/2009/01/02/wikipedia_fundraising_2m_jan_2/}
\BIBentrySTDinterwordspacing

\bibitem{reuters}
\BIBentryALTinterwordspacing
W.~Roelf. (2007) Wikipedia founder mulls revenue options. [Online]. Available:
  \url{http://www.reuters.com/article/internetNews/idUSL1964587420070420}
\BIBentrySTDinterwordspacing

\bibitem{wikifear}
\BIBentryALTinterwordspacing
H.~Leslie. (2007) Wikipedia to run out of money? [Online]. Available:
  \url{http://digital-lifestyles.info/2007/02/12/wikipedia-to-run-out-of-money/}
\BIBentrySTDinterwordspacing

\bibitem{vogel}
W.~Vogels, ``Eventually consistent,'' \emph{ACM Queue}, vol.~6, 2008.

\bibitem{hewitt2008osr}
C.~Hewitt, ``{ORG}s for scalable, robust, privacy-friendly client {C}loud
  {C}omputing,'' \emph{IEEE Internet Computing}, vol.~12, no.~5, 2008.

\end{thebibliography}

\end{document}